\documentclass[12pt]{iopart}
\usepackage{iopams}
\usepackage{graphicx}% Include figure files

\begin{document}

\title[Precision preparation of strings of trapped neutral atoms]
{Precision preparation of strings of trapped neutral atoms}

\author{Y~Miroshnychenko, W~Alt, I~Dotsenko, L~F\"{o}rster, M~Khudaverdyan, A~Rauschenbeutel and D~Meschede}

\address{Institut f\"{u}r Angewandte Physik, Universit\"{a}t Bonn, Wegelerstr. 8, D-53115 bonn, Germany}
\ead{meschede@iap.uni-bonn.de}
\begin{abstract}
We have recently demonstrated the creation of regular strings of
neutral caesium atoms in a standing wave optical dipole trap using
optical tweezers [Y. Miroshnychenko et al., Nature, in press
(2006)]. The rearrangement is realized atom-by-atom, extracting an
atom and re-inserting it at the desired position with sub-micrometer
resolution. We describe our experimental setup and present detailed
measurements as well as simple analytical models for the resolution
of the extraction process, for the precision of the insertion, and
for heating processes. We compare two different methods of
insertion, one of which permits the placement of two atoms into one
optical micropotential. The theoretical models largely explain our
experimental results and allow us to identify the main limiting
factors for the precision and efficiency of the manipulations.
Strategies for future improvements are discussed.

\end{abstract}

\maketitle

\section{Introduction}
Neutral atoms stored in light induced potentials form a versatile
tool for studying quantum many body systems with controlled
interactions. One of the most interesting cases occurs if the
coherence of such processes is preserved, and hence the build-up of
many body quantum correlations can be studied in detail. Such
experimental systems are of great interest for quantum information
processing \cite{ECRoadmap05}, and, more general, quantum simulation
\cite{Feynman82}. Using far detuned optical dipole traps, neutral
atoms can be well confined in various geometrical configurations,
while at the same time offering long coherence times of their
internal states \cite{Davidson95,Ozeri99}.

%Further applications of stored atoms include precision
%spectroscopy and frequency standards \cite{Katori05}.

%Interacting neutral atom many body systems can be in homogeneous
%large potentials as well as in periodic potentials with discrete
%sites can be realized. Here we concentrate on the latter, lattice
%type system.

%In all cases it is necessary to experimentally control the
%interactions of the particles with excellent accuracy.
Two general approaches towards the realization of suitable neutral
atom systems can be distinguished: In the typical "top-down"
approach one starts with a large sample of Bose-condensed atoms
which are then adiabatically transferred into a three-dimensional
optical lattice. A close to perfect array of 10$^5$ to 10$^6$
atoms is then obtained with almost exactly one atom per site by
inducing the Mott insulator state \cite{Greiner02}. For this
system, the method of spin dependent transport \cite{Mandel03} has
made possible the creation of large-scale entanglement by inducing
controlled, i.e. phase coherent, collisions between neighbouring
atoms. However, due to the small distance between adjacent atoms,
the manipulation and state detection of individual atoms is still
a big challenge.

This problem is overcome in our "bottom-up" approach where strings
of trapped neutral atoms are created one by one. We have
experimentally demonstrated \cite{Miroshnychenko06a} that regular
strings consisting of up to 7 atoms spaced several potential well
apart can be created in a one-dimensional optical lattice. Due to
the larger interatomic distance, we are able to address individual
atoms reliably \cite{Schrader04a}. Moreover, the exact number of
empty potential wells between two atoms has been measured
\cite{Dotsenko05}, enabling the method of spin dependent transport
in this system.

Small strings of neutral atoms are not only excellent experimental
objects to implement controlled coherent collisions, they are also
well suited for deterministic coupling using cavity-QED concepts
\cite{You03}. Here, atom-atom entanglement can be created by
synchronous interaction of two atoms with a single mode of the
cavity field. Typical modes have diameters of a few 10 $\mu$m,
compatible with the interatomic separations on the order of 5-15
$\mu$m in our strings.

In this manuscript we give a detailed analysis of the properties
and limitations of our atom sorting apparatus which we use to
create such regular strings.

\section{Experimental tools}
\subsection{Standing wave optical dipole traps}
\begin{figure}
\centering
\includegraphics{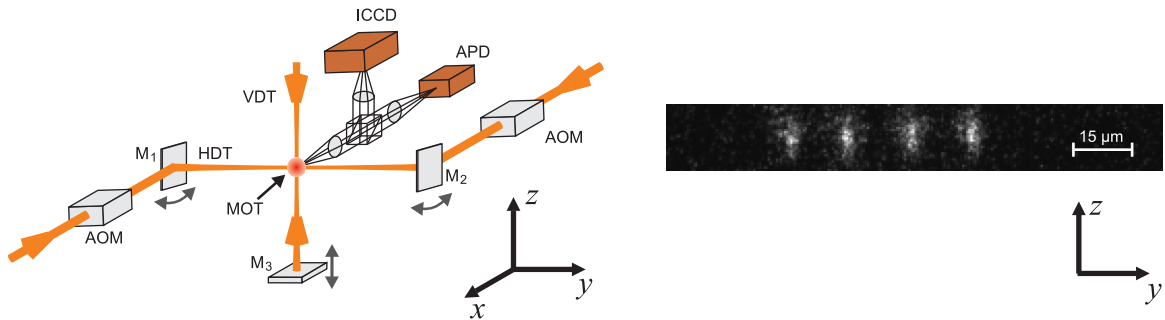}
\caption{\label{fig:setup} %(Color online)
Scheme of the experimental setup for the rearrangement of trapped
atoms. Two counter-propagating laser beams produce a horizontally
oriented optical standing wave dipole trap (HDT). After loading an
exact number of caesium atoms from a magneto-optical trap (MOT)
into the HDT, a fluorescence picture, recorded with an intensified
CCD camera (ICCD), reveals the initial positions of the atoms. A
second, vertically oriented standing wave dipole trap (VDT) is
used as optical tweezers to extract a selected atom out of the HDT
and to reinsert it at the desired position into the HDT. The ICCD
image shows four atoms which have been rearranged into a regular
string in the HDT (exposure time 1~s).}
\end{figure}

% Aim: Introduce dipole
%traps and their important parameters (wavelength (do not interfere),
%polarization, waist, depth). See
%Fig.~\ref{fig:setup}.\\
%
We trap neutral caesium atoms in a red-detuned optical standing wave
dipole trap, oriented horizontally (HDT) (see
figure~\ref{fig:setup}). It is formed by two counter-propagating
laser beams with parallel linear polarization, generating a chain of
potential well located at the intensity maxima. For a Nd:YAG laser,
the periodicity is $\lambda_\mathrm{HDT}/2=532$~nm. The beams are
focused to a waist radius of $w_\mathrm{HDT}=19~\mathrm{\mu m}$
yielding a Rayleigh range of 1~mm. An optical power of 1~W per beam
results in a measured trap depth of $U_\mathrm{HDT}^0=0.8$~mK.

Individual atoms trapped in the HDT can be extracted and
reinserted with another optical standing wave trap used as optical
tweezers. This trap (VDT) is oriented vertically and
perpendicularly to the HDT. The standing wave is created by retro
reflecting the linearly polarized beam of an Yb:YAG laser
($\lambda_\mathrm{VDT}=1030$~nm). This laser beam is focused to a
waist radius of $w_\mathrm{VDT}=10~\mathrm{\mu m}$. The typical
power of $0.3$~W creates a measured trap depth of
$U_\mathrm{VDT}^0=1.4$~mK. The power of the VDT laser beam is
controlled by an electro optical modulator (EOM).\\

%As the source of single atoms use MOT.

\subsection{Magneto-optical trap}
\label{subs:MOT}
%Aim: Introduce MOT parameters (high/low gradient and forced loading,
%escape energy).\\
%
Our vacuum chamber consists of a glass cell connected to an
ultra-high vacuum main chamber and a caesium reservoir separated
from the chamber by a valve. An ion getter pump maintains a
background gas pressure below $10^{-10}$~mbar.

We use a high gradient magneto-optical trap (MOT) as a source of
single atoms for our experiments \cite{Haubrich96}. The laser
system of the MOT consists of two diode lasers in Littrow
configuration, frequency-stabilized by polarization
spectroscopies. The cooling laser is stabilized to the
$F=4\rightarrow F'=3/F'=5$ crossover transition and shifted by an
acousto optic modulator (AOM) to the red side of the cooling
transition $F=4\rightarrow F'=5$. The $z$-axis of the MOT
coincides with the axis of the VDT, whereas the two other axes of
the MOT are in the $x$-$y$-plane at $45^0$ to the axis of the HDT.
The saturation parameter $s=\frac{I}{I_0} \left[ 1+\left( \frac{2
\Delta}{\Gamma}\right)^2\right]^{-1}$ of each MOT beam is $s=0.5$
($I$: intensity of the cooling laser; $I_0=1.1~\mathrm{mW/cm^2}$:
saturation intensity of the caesium D2 transition;
$\Gamma=2\pi\cdot 5.2$~MHz: linewidth of the excited state 6P$_3
/2$; $\Delta=1.5~\Gamma$: detuning of the cooling laser).
%The saturation parameter $s=\pi h c/3\lambda^2\tau$ ($\lambda$ =
%852~nm: Resonant wave length of the caesium D2 transition; $\tau$ =
%30~ns: Lifetime of excited 6P$_{3/2}$ state) of each MOT beam is
%$s=0.5$.
The MOT repumping laser is stabilized to the $F=3\rightarrow F'=4$
transition. It is linearly polarized and propagates along the axis
of the HDT. We typically use $1~\mathrm{\mu W}$ focused to about
$0.5$~mm.

The high magnetic field gradient of the MOT ($\partial B/\partial
z=340$~G/cm) is produced by water cooled magnetic coils mounted
symmetrically with respect to the glass vacuum cell. The magnetic
field can be switched within 60~ms (mainly limited by the eddy
currents in the metal parts of the coils). Due to the high field
gradient, the spontaneous loading rate of Caesium atoms from the
thermal background vapor into the MOT is negligibly slow.

In order to load the MOT, we temporarily reduce the magnetic field
gradient to $\partial B/\partial z=25$~G/cm for the time
$\tau_\mathrm{load}$ to increase the capture cross section.
Varying the loading time $\tau_\mathrm{load}$ from few tens to few
hundred milliseconds, we can select a specific average number of
loaded atoms ranging from 1 to 50.

The atoms are transferred from the MOT into the HDT by
simultaneously operating both traps for several tens of
milliseconds.

\begin{figure}
\centering
\includegraphics{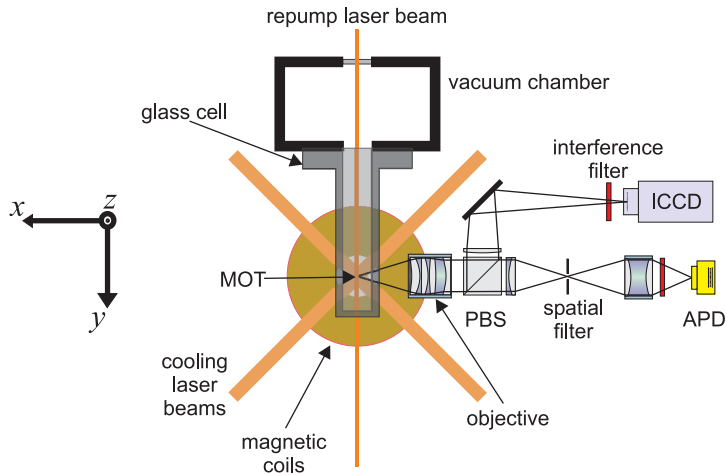}
\caption{\label{fig:detection_optcs} Experimental setup with the
detection optics for the MOT fluorescence. The fluorescence light is
collected and collimated by an objective. One part of the
fluorescence signal is spatially and spectrally filtered and
detected with an avalanche photo diode (APD). The other part is only
spectrally filtered and sent into an intensified CCD camera.}
\end{figure}

%Atoms can be counted and imaged.

\subsection{Atom detection}
%Aim: Introduce Imaging optics, APD, ICCD. Fig.~\ref{fig:setup}.\\
%
The procedures described in this paper rely on our ability to
nondestructively determine the exact number and the position of
trapped atoms by detecting their fluorescence. For this purpose
the fluorescence light is collected by a homemade long working
distance microscopic objective (NA=0.29), covering about $2~\%$ of
the solid angle \cite{Alt02}.
%[Alt's Objective paper].
The fluorescence is monitored in the time domain by an avalanche
photo diode (APD, type SPCM200 CD2027 from EG$\&$G, quantum
efficiency $50~\%$ at 852~nm) and spatially by an intensified CCD
camera (ICCD, type PI-MAX:1K,HQ,RB from Princeton Instruments with
image intensifier Gen III HQ from Roper Scientific, quantum
efficiency $10~\%$ at 852~nm) \cite{Miroshnychenko03},
%[Optics express]
see figure~\ref{fig:detection_optcs}.

\subsubsection{Atom number detection}
%
%Aim: Optimal integration time.\\
%

\begin{figure}
\centering
\includegraphics[width=8cm]{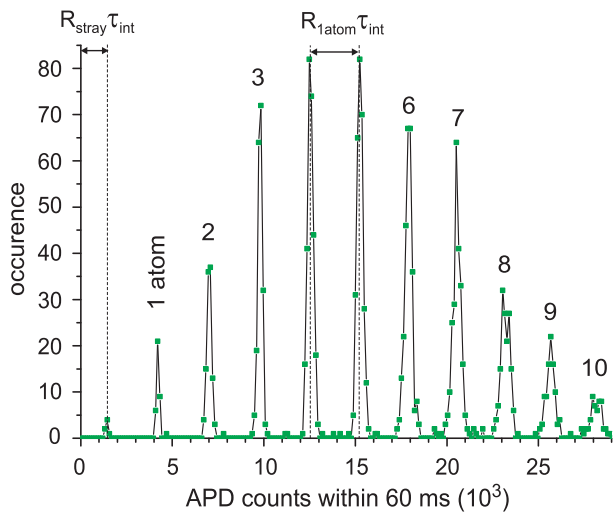}
\caption{\label{fig:fluorescenceRate} Histogram of the APD counts
detected within the integration time
$\tau_\mathrm{int}=60~\mathrm{ms}$ after loading five atoms on
average into the MOT (1800 repetitions). The peaks correspond to
different number of atoms in the MOT.}
\end{figure}

Our atom counting method exploits the fact that each atom
contributes equally to the intensity of the MOT fluorescence signal
and on the high signal to noise ratio of our detection system,
allowing us to distinguish discrete levels in the APD count rate.
For $n$ trapped atoms we detect $N_n=(R_\mathrm{stray}+n\cdot
R_\mathrm{1atom})\cdot \tau_\mathrm{int}$ photons during the
integration time $\tau_\mathrm{int}$, see
figure~\ref{fig:fluorescenceRate}. Here, $R_\mathrm{stray}$ is the
count rate due to the stray light and the detector background, and
$R_\mathrm{1atom}$ is the actual one-atom fluorescence rate. For
typical MOT parameters
% $50~\%$
%loss on the beamsplitter,
we detect $R_\mathrm{1atom}=35000/\mathrm{s}$ and
$R_\mathrm{stray}=25000/\mathrm{s}$.

%Synchronously, the CCD camera receives about
%$R_\mathrm{1atom}^\mathrm{CCD}= 7000/\mathrm{s}$.

The standard deviation of the detected photon number is
fundamentally limited by Poisson statistics to $N_n^{-1/2}$.
Fluctuations of the MOT laser beams including intensity, phase and
pointing stability are taken into account in analogy with the
description of intensity noise in laser beams by a global relative
intensity noise $RIN = \delta N^2/N^2$ where $\delta N^2$
represents the rms-value of these fluctuations. In our case $RIN =
0.01^2-0.02^2$.

In order to distinguish between $n$ and $n+1$ atoms, the total
width of the peaks corresponding to neighboring atom numbers
($\Delta N= (N_n+RIN\cdot N_n^2)^{1/2}$) has to be compared to
their separation $R_\mathrm{1atom}\tau_\mathrm{int}$. In order to
distinguish atom numbers with better than 95\% confidence, the
ratio
\begin{equation}
k_n = \frac{\Delta N}{N_{n+1}-N_n} = \left( \frac{R_\mathrm
{stray}}{R_\mathrm{1atom}}+ n \right) \sqrt{ \frac{1}{(R_\mathrm
{stray} + n R_{\mathrm{1atom}})\tau_\mathrm{int}}+RIN }
%k_n=N_n^{-1/2}/ (N_{n+1}-N_n) \approx \sqrt{n/(\tau_\mathrm{int}
%R_\mathrm{1atom})}
\label{eq:confidenceRatio}
\end{equation}
must be smaller than $1/4$. In our experiments, RIN begins to
dominate this ratio for integration times $\tau_\mathrm{ int}
\simeq 60~\mathrm{ms}$. This time was chosen discriminating atom
numbers from the APD signal, since it is short compared to other
experimental procedures, and longer times do not improve the
signal to noise ratio. This method allows us to discriminate 1 to
20 atoms in the MOT with a confidence level above $95~\%$.

\subsubsection{Atom position detection}
%Aim: Calibration $\mu m$-pixel Fig.~\ref{fig:calibration}, precision
%of the position determination.
%

The positions of the atoms in the HDT are determined by illuminating
them with an optical molasses and detecting the fluorescence with
the ICCD camera. At the same time, the optical molasses cools the
trapped atoms, enabling continuous observations of up to one
minute~\cite{Miroshnychenko03}. We detect about 160 photons per atom
on the ICCD camera within the 1~s exposure time. The $y$-positions
of the individual atoms trapped in the HDT (see
figure~\ref{fig:setup}) are determined by binning the pixels of the
ICCD image in the vertical $z$-direction after suitably clipping the
image to minimize background noise. The resulting one-dimensional
intensity distribution along the $y$-direction is fitted with a sum
of Gaussians, which are used as an approximation to the line spread
function of our imaging system \cite{Dotsenko05}. We define the
centers of the Gaussians as the $y$-positions of the atoms, which
can be determined with a precision of 140~nm~rms (below the
wavelength of the imaging light) within 1500~ms (1000~ms of exposure
and 500~ms read-out and image processing). In this way we are able
to determine the
number of potential wells separating two simultaneously trapped atoms. \\

Since we want to transport atoms over distances up to 1~mm with
submicrometer accuracy, it is essential to obtain a precise
calibration of camera pixel to the position in the object plane of
the microscope objective.

\begin{figure}
\centering
\includegraphics[width=14cm]{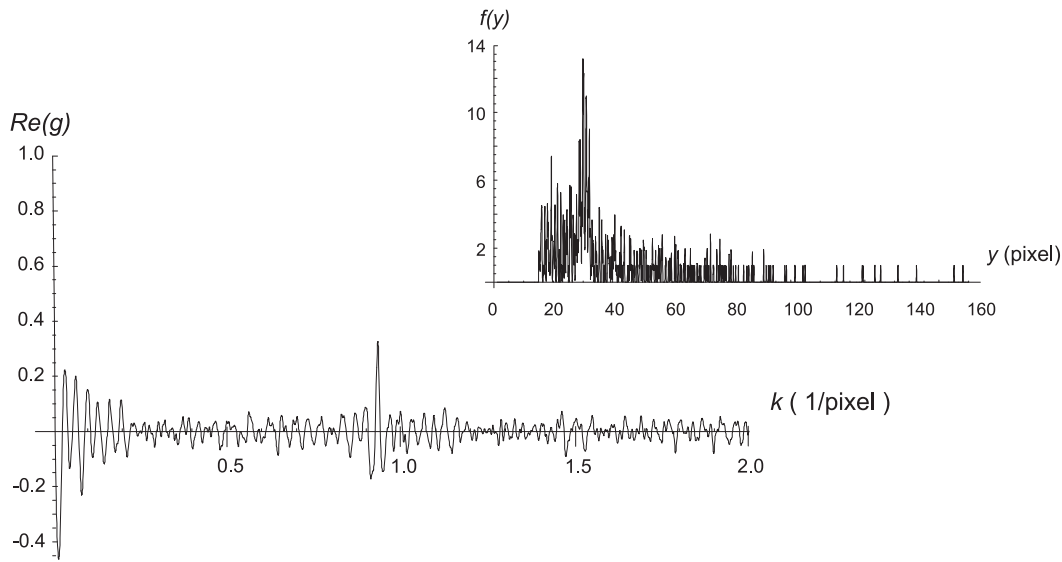}
\caption{\label{fig:calibration} Histogram of measured distances and
its fourier transfor. The only one prominent contribution at
$k_0=0.9336~\mathrm{wells/pixel}$ corresponds to the periodicity of
$\lambda_\mathrm{HDT}/2$ in the object plane of the microscope
objective. The inset shows the original function $f(y)$. Here, for
presentation purposes each delta function was replaced by a Gaussian
with the width of $0.05$~pixel.}
\end{figure}

For this calibration we take advantage of the fact that the atoms in
the HDT are trapped in the potential minima separated by exactly
$\lambda_\mathrm{HDT}/2=532$~nm. Therefore, the measured distance
between two simultaneously trapped atoms $d$ given in units of
camera pixels must correspond to an integer multiple of
$\lambda_\mathrm{HDT}/2$ in the object plane:
\begin{equation}
\alpha d =n \lambda_\mathrm{HDT}/2,
 \label{eq:calibration}
\end{equation}
where $\alpha$ is the calibration parameter in $\mathrm{\mu
m/pixel}$. In order to determine $\alpha$ we have first accumulated
about 500 images with two to four atoms trapped in the HDT. Then we
have determined the interatomic separations in each image, resulting
in $n\approx700$ distance values $d_i$, shown in
figure~\ref{fig:calibration}. In order to avoid any inaccuracy
caused by overlapping peaks at short distances, only separations of
more than $10~\mathrm{\mu m}$ then were taken into account. To find
the periodicity of the distribution we construct a function built by
summing the delta functions at the positions of each $d_i$
\begin{equation}
f(y)=\frac{1}{n}\sum_{i=1}^n \delta(d_i-y)
 \label{eq:periodicFunction}
\end{equation}
and Fourier transform it:
\begin{equation}
g(k)=\frac{1}{\sqrt{2\pi}}\int_{-\infty}^{\infty} f(y) e^{2\pi i k
y}d y=\frac{1}{\sqrt{2\pi}~ n}\sum_{i=1}^n e^{2\pi i k d_i}.
 \label{eq:fourierImage}
\end{equation}
The real part of the Fourier transform of $g(k)$ is shown in
figure~\ref{fig:calibration}. The most prominent peak at
$k_0=0.9336(\pm 0.0003)~1/\mathrm{pixel}$ corresponds to the spatial
frequency of the standing wave pattern:
\begin{equation}
k=\frac{\alpha}{\lambda_\mathrm{HDT}/2}~~.
 \label{eq:spatialFreq}
\end{equation}
This yields the calibration parameter $\alpha=0.4967(\pm
0.0002)~\mathrm{\mu m/pixel}$.

The error of this value is dominated by the statistical error due to
the finite sample and the $130~\mathrm{nm}$-uncertainty in the
determination of each distance \cite{Dotsenko05}. The statistical
error is estimated by randomly selecting a subset of $n/2$ distances
and determining $\alpha$ by the above mentioned calculations on this
subset. Using 20 different subsets the standard deviation $(\delta
\alpha)_{n/2}$ was determined. The statistical error for the full
set is therefore $(\delta \alpha)_{n}=(\delta
\alpha)_{n/2}/\sqrt{2}$. The slight modification of the wave length
$\lambda_\mathrm{HDT}$ in the Rayleigh zone by the Guoy phase is on
the order of 10$^{-5}$ and hence negligible here.

\subsection{Three-dimensional transport of atoms}
%Aim: Methods to transport (along $x$-, $y$- and $z$-directions),
%timescales, precision.\\
%
We transport atoms in the $x$-$y$-plane using the HDT. Vertical
transport of the atoms along the $z$-direction is realized by the
VDT.

\subsubsection{Transportation along the $y$-direction}
An ``optical conveyor belt'' \cite{Kuhr01} along the trap axis is
realized by means of acousto-optic modulators (AOMs) installed in
each arm of the HDT, see figure~\ref{fig:setup}. Mutually detuning
the AOM driving frequencies using a dual-frequency synthesizer,
detunes the frequencies of the two laser beams. As a result, the
standing wave pattern moves along the axis of the trap.

The optical conveyor belt allows us to transport the atoms over
millimeter distances with submicrometer precision
\cite{Dotsenko05,Kuhr01} within several milliseconds. The accuracy
of the transportation distance is limited to $190~\mathrm{nm}$ by
the discretisation error of our digital AOM-driver
\cite{Dotsenko05}. In this experiment we typically transport atoms
over a few tens of micrometers within a few hundred microseconds.

\subsubsection{Transportation along the $x$-direction}
Displacement of the HDT along the $x$-direction is realized by
synchronously tilting the mirrors M$_1$ and M$_2$, see
figure~\ref{fig:setup}, in opposite directions around the $z$-axis
using piezo-electric actuators. For tilt angles below
$0.1~\mathrm{mrad}$ the variation of the interference pattern is
small and to a good approximation pure $x$-translation is realized.

We typically move atoms in the $x$-direction by two times the waist
radius of the HDT (ca. $40~\mathrm{\mu m}$) with a precision of a
few micrometers within 50~ms. The maximum transportation distance is
limited to about $40~\mathrm{\mu m}$ by the dynamic range of the
actuators. The minimal transportation time is limited to about
$10~\mathrm{ms}$ by the bandwidth of the PZT-system.

\subsubsection{Transportation along the $z$-direction}
The VDT acts as optical tweezers and extracts and reinserts atoms in
the $z$-direction. To axially move the standing wave pattern of the
VDT, the retro-reflecting mirror $M_3$ is mounted on a linear PZT
stage, see figure~\ref{fig:setup}.

In our experiments the VDT transports an atom over about
$70~\mathrm{\mu m}$ along the $z$-axis by applying a sinusoidal
voltage ramp to the PZT within $50~\mathrm{ms}$. The precision of
the transportation is limited to a few micrometers by the hysteresis
of the piezo-crystal, whereas the transportation time is limited by
the inertia of the mirror.

\section{Positioning individual atoms in the HDT}
\subsection{Outline of the distance-contlor procedure}
\label{subs:repositioning_procedure}
%Aim: Introduce the feedback algorithm: load, observe, extract,
%re-insert.
\begin{figure}
\centering
\includegraphics[width=8cm]{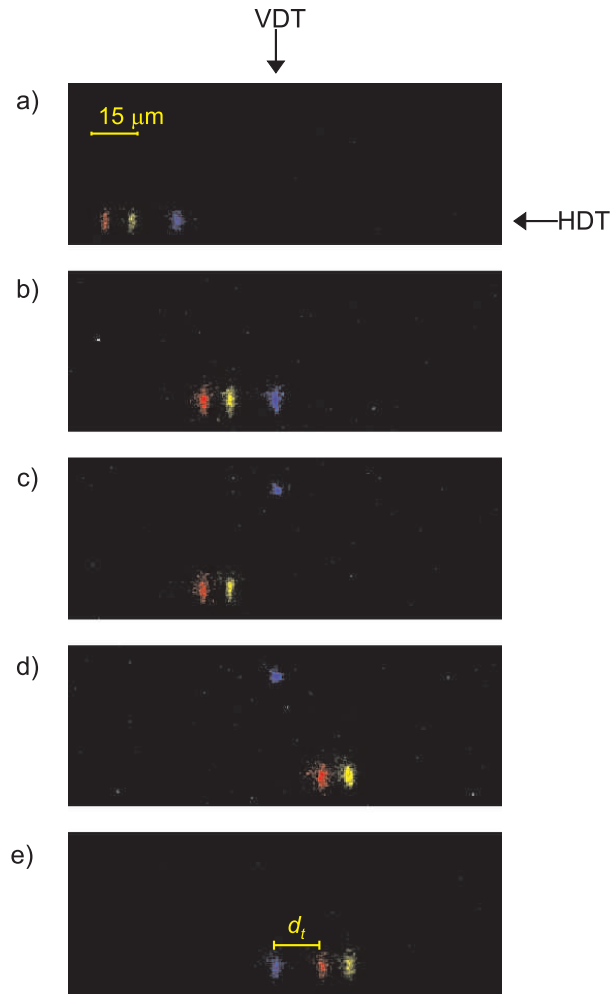}
\caption{\label{fig:sortingProcedure}Distance-control operation
\cite{Miroshnychenko06a}. After the initial positions of all the
atoms are determined (a), the rightmost atom is transported to the
position of the VDT (b). This atom is extracted vertically with the
optical tweezers out of the HDT (c). The rest of the atoms is
transported along the axis of the horizontal trap, so that the
leftmost atom arrives at the target separation $d_t$ from the axis
of the VDT (d). The extracted atom is inserted at the desired
position into the HDT (e). Note that the fluorescence spots,
corresponding to respective atoms, were colored for visualization
purposes.}
\end{figure}

%-> position within the string

Immediately after loading the HDT with $n\geq 2$ atoms, they are
randomly distributed over an interval of about $100~\mathrm{\mu m}$
along the axis of the trap. In order to create regular strings with
a target interatomic separation $d_\mathrm{t}$, atoms are
repositioned one by one with the VDT-optical tweezers. For this, the
initial positions of all atoms are first determined by recording and
analyzing a fluorescence ICCD image, see
figure~\ref{fig:sortingProcedure}(a). Then, the atoms are rearranged
in the HDT by sequential application of the ``distance-control''
operation: The string of atoms in the HDT is transported
horizontally along the trap axis, such that the rightmost atom
arrives at the $y$-position of the VDT ($y_\mathrm{VDT}$), see
figure~\ref{fig:sortingProcedure}(b). After adiabatically switching
on the VDT, this atom is transported upwards by approximately three
times the waist of the HDT, out of its region of influence.

\begin{figure}
\centering
\includegraphics[width=14.8cm]{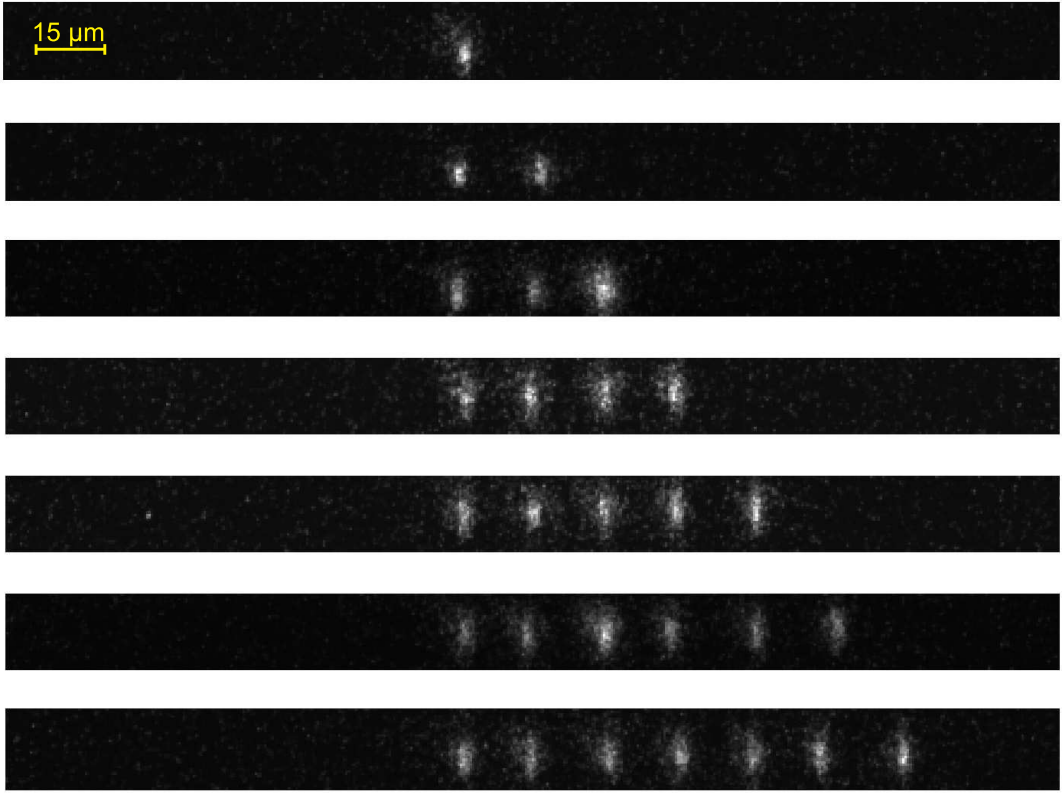}
\caption{\label{fig:strings}Equidistant strings of $n$ atoms created
by rearranging the atoms of the string applying $n-1$ times the
distance-control operation.}
\end{figure}

This atom is then extracted with the optical tweezers (c). The rest
of the string in the HDT is transported along the $y$-axis until the
leftmost atom of the string arrives at $y_\mathrm{VDT}+d_\mathrm{t}$
(d). The procedure is completed by reinserting the extracted atom at
this position into the HDT (e). Each operation permutes the order of
the atoms, and after $n-1$ steps an equidistant string of $n$ atoms
is formed. Figure~\ref{fig:strings} shows ICCD images of equidistant
strings of up to seven atoms with interatomic
separations of $15~\mathrm{\mu m}$.\\

%Further in this section, we first analyze the two key processes: the
%extraction (Sec.~\ref{subs:extraction}) and of the re-insertion
%(Sec.~\ref{subs:insertion}) of an atom into the HDT using the
%optical tweezers. Then, the precision (Sec.~\ref{subs:precision})
%and the efficiency (Sec.~\ref{subs:efficiency}) of the distance
%control operation is analyzed.

A perfect distance-control procedure would extract and re-insert an
atom with $100~\%$ efficiency at a separation to the next atom given
in terms of an exactly known number of micropotentials always, i.
e., multiple of $\lambda_\mathrm{HDT}/2$. We have developed a model
of extraction and re-insertion in order to study the physical
limitations of the repositioning procedure.

%We find good agreement of the models and experimental data. Spatial
%fluctuations and drifts of the VDT with respect to the HDT and the
%temperature of the trapped atoms are the main factors limiting the
%precision.

\subsection{Extraction of an atom}
\label{subs:extraction}
%\subsubsection{Principle of operation}
%Aim: Introduce the physical processes necessary for extraction.\\

For extraction, the VDT-optical tweezers needs to overcome the HDT
trapping forces. In both the HDT and VDT standing wave dipole traps
confinement in the axial direction is almost two orders of magnitude
tighter than in the radial direction, the maximal axial forces are
thus much larger than the radial forces. For comparable potential
depths of the HDT and the VDT, an atom in the overlap region will
therefore always follow the axial shift of the traps.

%standing waves, here the VDT in the $z$-direction.

Successful extraction of a single atom not only requires efficient
handling of the atoms between the HDT and VDT traps. In addition,
other atoms present in the vicinity must remain undisturbed. We have
thus defined and analyzed a minimal separation of atoms tolerable on
extraction, which is equivalent to an effective ``width of the
optical tweezers''.

%Important parameter: ''tweezers size`` - how close the atoms can be
%to separately extract one of them.

\subsubsection{Theoretical model of the width of the optical tweezers}
%Aim: Introduce the theory, see Fig.~\ref{fig:potentials}.
In this model motion in the traps is treated classically, for at the
atomic temperature of about $60~\mathrm{\mu K}$ for the typical
depths of the traps in our experiments the mean oscillatory quantum
numbers are $n_\mathrm{rad}\approx 90$, $n_\mathrm{ax}\approx 3$ for
the VDT, and $n_\mathrm{rad}\approx 400$, $n_\mathrm{ax}\approx 6$
for the HDT.

We consider two crossed standing wave optical dipole traps. For
simplicity, we assume that all the spatial manipulations are carried
out within the Rayleigh-range of the standing wave dipole traps, i.
e., we neglect the change of the curvature of the wave fronts. In
this approximation, each dipole trap is described by three
parameters: the waist radius of the Gaussian beam, the depth of the
trap, and the periodicity of the standing wave. Atoms are trapped in
the different potential wells of the standing wave of the HDT and
are extracted purely along the $z$-direction. The motion occurs in
the $y$-$z$-plane only. Therefore, we consider one-dimensional
potentials along the $z$-axis at different $y$-positions in the
$y$-$z$-plane with $x=0$.
\begin{figure}
\centering
\includegraphics[width=14.8cm]{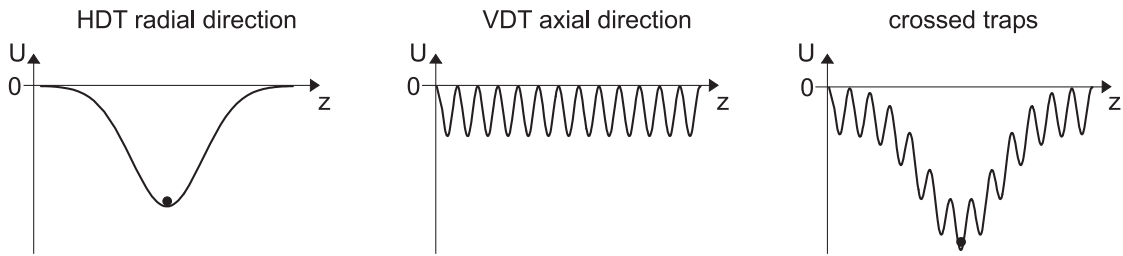}
\caption{\label{fig:potentials}Trapping potentials along the
$z$-axis. The trapping potential in the overlap region of the two
traps is the sum of the Gaussian profile of the radial confining
potential of the HDT and of the standing wave of the VDT. An atom at
the bottom of the HDT experiences light forces of the standing wave,
too.}
\end{figure}

In order to separate the effects of the potential shape and the
atomic motion on the process of extraction, we first model the case
of atoms at zero temperature, where the energy of the atoms is well
defined. Later, the influence of the thermal
energy distribution is discussed.\\
\\
\textit{Tweezers potential} Consider an atom at rest trapped at the
bottom of a micropotential of the HDT at $y=0$, which coincides with
the axis of the VDT. At this position the HDT-potential in the
$z$-direction has a Gaussian shape with waist $w_\mathrm{HDT}$ and
depth $U_\mathrm{HDT}^0$. After switching on the VDT, in addition to
the Gaussian potential of the HDT, a periodic potential with depth
$U_\mathrm{VDT}^0$ and period $\lambda_\mathrm{VDT}/2$ is
superimposed in the $z$-direction, see figure~\ref{fig:potentials}.
Since the HDT and VDT laser frequencies are far apart, the trapping
potentials are added incoherently, and the atom is then additionally
subject to the forces of the VDT standing wave.

\begin{figure}
\centering
\includegraphics[width=13cm]{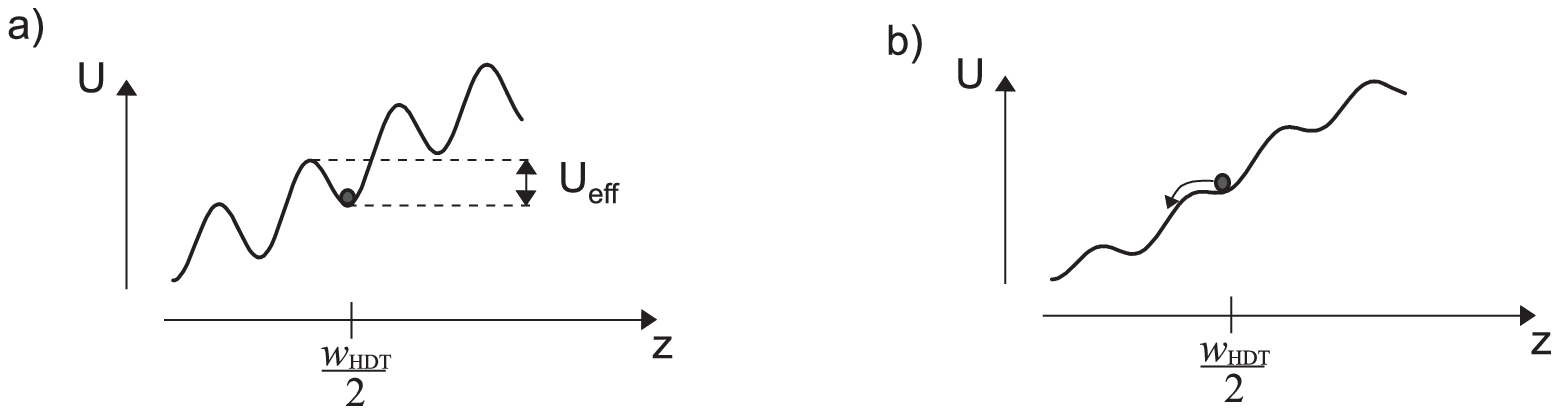}
\caption{\label{fig:slope}Decrease of the effective depth of the
potential well. Due to the Gaussian shape of the confining potential
of the HDT in the $z$-direction, the depth of the standing wave
pattern reaches its minimum at the distance $w_\mathrm{HDT}/2$ from
the axis of the HDT. If the effective depth $U_\mathrm{eff}$ is
greater than zero, the atom remains trapped in the standing wave and
will be finally extracted with the VDT a). Otherwise, the atom
always rolls down during the extraction b).}
\end{figure}

During extraction of the atom during the $z$-direction it is
conveyed at the bottom of the micropotential away from the axis of
the HDT. Due to the Gaussian radial profile of the HDT, the depth of
the local potential minima changes along the $z$-axis and reaches
its minimum at the distance $z_\mathrm{max}=w_\mathrm{HDT}/2$ from
the axis of the HDT. Here the slope of the Gaussian is maximal and
the effective depth of the local micropotential, see
figure~\ref{fig:slope}, can be approximated as
\begin{equation}
U_\mathrm{eff}\approx U_\mathrm{VDT}^0- \frac{1}{\pi
\sqrt{e}}\frac{\lambda_\mathrm{VDT}}{w_\mathrm{HDT}}U_\mathrm{HDT}^0.
 \label{eq:uVDTmodified}
\end{equation}
The condition for extracting the atom from the HDT is given by
\begin{equation}
U_\mathrm{eff}>0.
 \label{eq:uCondition}
\end{equation}
%Hence, the lower limit for the VDT potential is
%\begin{equation}
%U_\mathrm{VDT}^0>
%\frac{1}{2\sqrt{e}}\frac{\lambda_\mathrm{VDT}}{w_\mathrm{HDT}}U_\mathrm{HDT}^0.
%%U_\mathrm{VDT}^0> \frac{\lambda_\mathrm{VDT}}{4}S_\mathrm{HDT}.
% \label{eq:uVDTcondition}
%\end{equation}

\begin{figure}
\centering
\includegraphics[width=8cm]{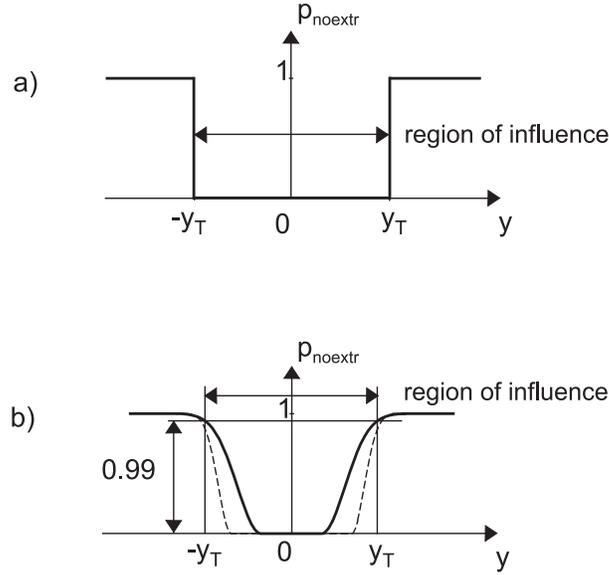}
\caption{\label{fig:ExtractionProbability}Probability for an atom
not to be extracted from the HDT. a) atoms at $T=0~\mathrm{K}$. b)
atoms at temperatures $T_1$ (solid line) and $T_2$ (dashed line),
with $T_1>T_2$.}
\end{figure}

Now, consider an atom is trapped at some other position $y\neq0$
along the HDT. The potential along the $z$-direction will be the sum
of the same Gaussian potential well of the HDT with depth
$U^0_\mathrm{HDT}$ and of the periodic potential of the VDT, but now
with the reduced depth $U_\mathrm{VDT}^0~e^{-\frac{2
y^2}{w_\mathrm{VDT}^2}}$. The sum of the two potentials at the
lateral position $y$ generalizes (\ref{eq:uVDTmodified}) to
\begin{equation}
U_\mathrm{eff}(y) \approx
U_\mathrm{VDT}^0~e^{-\frac{2y^2}{w_\mathrm{VDT}^2}}-\frac{1}{\pi\sqrt{e}}\frac{\lambda_\mathrm{VDT}}{w_\mathrm{HDT}}U_\mathrm{HDT}^0.
%U(y)=U_\mathrm{VDT}^0~e^{-\frac{2y^2}{w_\mathrm{VDT}^2}}-\frac{\lambda_\mathrm{VDT}}{4}S_\mathrm{HDT}.
 \label{eq:uVDTmodified2}
\end{equation}
Consequently, there exists some region
$-y_\mathrm{T}<y<y_\mathrm{T}$ along the HDT, where condition
(\ref{eq:uCondition}) holds, and where atoms will be extracted by
the VDT. Figure~\ref{fig:ExtractionProbability}(a) shows the
probability $p_\mathrm{noextr}$ for an atom to remain trapped in the
HDT after the extraction as the function of the lateral position
$y$. The critical position $y_\mathrm{T}$ defined by the condition
$U_\mathrm{eff}(y_\mathrm{T})=0$ is
\begin{equation}
%y_\mathrm{T}=\frac{w_\mathrm{VDT}}{\sqrt{2}}
%\sqrt{\ln\left(U_\mathrm{VDT}^0/\left(\frac{\lambda_\mathrm{VDT}}{4}S_\mathrm{HDT}\right)\right)}~~,
y_\mathrm{T}=\frac{w_\mathrm{VDT}}{\sqrt{2}} \sqrt{\ln\left(\pi
\sqrt{e}\frac{w_\mathrm{HDT}}{\lambda_\mathrm{VDT}}\frac{U_\mathrm{VDT}^0}{U_\mathrm{HDT}^0}\right)}~~.
 \label{eq:yCritical}
\end{equation}
%is
%\begin{equation}
%y_\mathrm{T}=\frac{w_\mathrm{VDT}}{\sqrt{2}}
%\sqrt{\ln\left(U_\mathrm{VDT}^0/\left(\frac{\lambda_\mathrm{VDT}}{4}S\right)\right)}~~,
% \label{eq:yCritical}
%\end{equation}
This equation characterizes the width of the optical tweezers,
$y_\mathrm{T}$, as a function of the trap parameters for atoms at
zero temperature. It shows that the lowering of $U_\mathrm{VDT}^0$
reduces the extraction width $2 y_\mathrm{T}$, from which the atoms
will be extracted. For $T=0~\mathrm{K}$ and neglecting quantum
effects, this region can be made arbitrary small at
$U_\mathrm{VDT}^0=
\frac{\lambda_\mathrm{VDT}}{\pi \sqrt{e}w_\mathrm{HDT}}U_\mathrm{HDT}^0$.\\
\\
%$U_\mathrm{VDT}^0=
%\frac{\lambda_\mathrm{VDT}}{4}S_\mathrm{HDT}$.
\textit{Thermal atomic motion} We now model atomic motion in the
dipole trap by an ensemble in thermal equilibrium at temperature $T$
in a three-dimensional harmonic potential. We assume that the energy
of the atoms is Boltzmann-distributed \cite{Bagnato87}:
\begin{equation}
f(E,T)=\frac{1}{2(k_\mathrm{B} T)^3}E^2 e^{-E/(k_\mathrm{B}T)}.
 \label{eq:maxwell_boltzmann}
\end{equation}
For an atom with a fixed energy $E$ the condition for the extraction
analogous to (\ref{eq:uCondition}) is
\begin{equation}
U_\mathrm{eff}(y)-E>0~.
 \label{eq:uCondition2}
\end{equation}
For a given temperature $T$ the fraction of atoms with an energy
above $U_\mathrm{eff}$ is given by
\begin{equation}
p(U)=\int_{\max{\{U_\mathrm{eff},0\}}}^\infty f(E,T)dE~,
 \label{eq:pU}
\end{equation}
which therefore is the fraction $p_\mathrm{noextr}$ of the atoms not
extracted from the HDT. As a function of the lateral position $y$ we
have
\begin{equation}
p_\mathrm{noextr}(y)\equiv p(U_\mathrm{eff}(y))=\frac{1}{2} \left[
\left(~\frac{U_\mathrm{eff}(y)}{k_\mathrm{B}T}+1\right)^2 +1
\right]\exp\left({-\frac{U_\mathrm{eff}(y)}{k_\mathrm{B}T}}\right).
 \label{eq:pHDT}
\end{equation}

Figure~\ref{fig:ExtractionProbability}(b) shows $p_\mathrm{noextr}$
for the same trap parameters as in
figure~\ref{fig:ExtractionProbability}(a). Atomic motion causes
``softening'' of the edges of the extraction zone. An increasing
temperature causes narrowing of the region of efficient extraction.

Here we define the region influenced by the optical tweezers $[
-y_\mathrm{T},y_\mathrm{T}]$, see
figure~\ref{fig:ExtractionProbability}(b), by
\begin{equation}
p_\mathrm{noextr}(y_\mathrm{T})\leq 0.99.
\end{equation}
In order to optimize the extraction resolution we vary
$U_\mathrm{VDT}^0$ such that $y_\mathrm{T}$ is minimal while still
guarantying efficient extraction in the center. We therefore choose
$U_\mathrm{VDT}^0$ fulfilling the condition
\begin{equation}
p_\mathrm{noextr}(y=0)=0.01
 \label{eq:p01}
\end{equation}
and get the minimal region of influence $[
-y_\mathrm{T}^\mathrm{opt},y_\mathrm{T}^\mathrm{opt}]$ from
\begin{equation}
p_\mathrm{noextr}(y_\mathrm{T}^\mathrm{opt})=0.99.
 \label{eq:p99}
\end{equation}
\begin{figure}
\centering
\includegraphics[width=15cm]{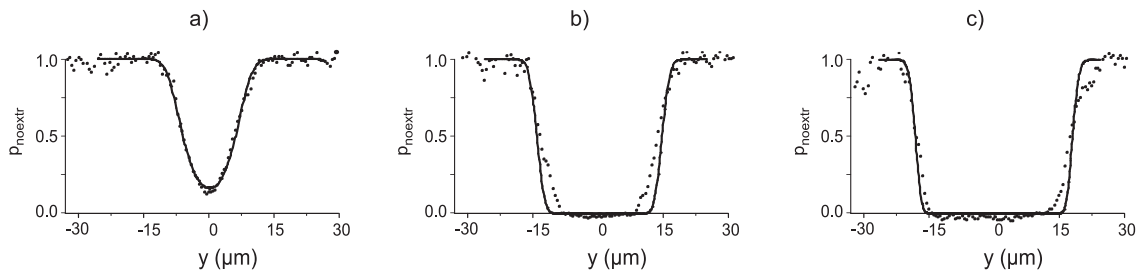}
\caption{\label{fig:extractionFits}Comparison of the experimental
data and the respective theoretical expectation. a) Fit of
$P_\mathrm{HDT}$ (solid line) to the experimental data (points) for
$U^0_\mathrm{VDT}/k_\mathrm{B}=0.3$~mK. b,c) The function
$P_\mathrm{HDT}$ with parameters from a) except
$U_\mathrm{VDT}^0/k_\mathrm{B}=3.1$~mK and
$U_\mathrm{VDT}^0/k_\mathrm{B}=16.8$~mK, respectively. The depth of
the HDT is $U_\mathrm{HDT}^0/k_\mathrm{B}=0.8$~mK.}
\end{figure}

% To verify the theory made
%an experiment.

\subsubsection{Measurement of the width of the optical tweezers}
%Aim: Experimental parameters and results.
We have experimentally determined the width $2y_\mathrm{T}$ of the
optical tweezers as a function of the depth of the VDT by loading
the HDT with atoms distributed over a region larger than
$y_\mathrm{T}$, extracting atoms with the VDT, and analyzing the
distribution of the atoms remaining in the HDT. Images of the atoms
in the HDT were taken before and after extraction, and used to
calculate the probability $P_\mathrm{HDT}(y)$ of the atoms to remain
trapped in the HDT after the extraction. In this measurement, the
depth of the HDT was fixed ($U_\mathrm{HDT}^0/k_\mathrm{B}=0.8$~mK),
whereas the depth of the VDT was varied over two orders of magnitude
from 0.3 to 16.8~mK. The corresponding plots for
$U^0_\mathrm{VDT}/k_\mathrm{B}=0.3$~mK, $3.1$~mK and $16.8$~mK are
presented in figure~\ref{fig:extractionFits}.

\subsubsection{Analysis}
%Aim: Fit the theory to the experiment, see
%Fig.\ref{fig:extractionFits}. Discuss the limiting factors.
In figure~\ref{fig:extractionFits}, the measured data are compared
to the theoretical model described by (\ref{eq:pHDT}). Free fit
parameters for the data of figure~\ref{fig:extractionFits}(a)
include the temperature $T$ of the atoms, the waist of the VDT
$w_\mathrm{VDT}$ along the axis of the HDT, and the position of the
VDT $y_0$, relative to the picture. The fit to the data set for the
depth of the VDT at $U^0_\mathrm{VDT}/k_B=0.3$~mK, corresponding to
a power of the incoming VDT laser beam of 0.06~W, yields
$$T=60(\pm 1)~\mathrm{\mu K}~~~\mathrm{and}~~~w_\mathrm{VDT}=11.6(\pm 0.2)~\mathrm{\mu m},$$
%$$y_0=-0.06(\pm 0.07)~\mathrm{\mu m}, $$
see figure~\ref{fig:extractionFits}(a). The temperature thus
obtained is in the range of the typical temperatures measured by
other methods \cite{Alt03}. Whereas the error is probably too small
and underestimate the systematic influence of the approximations of
the model. Also, the fitted value of the waist of the VDT is in
reasonable agreement with the value of $10.1(\pm 1.4)~\mathrm{\mu
m}$ determined from the oscillation frequency measurements. In
figure~\ref{fig:extractionFits}(b) and (c) we have plotted the model
function $p_\mathrm{noextr}(y)$ at
$U_\mathrm{VDT}^0/k_\mathrm{B}=3.1$~mK and
$U_\mathrm{VDT}^0/k_\mathrm{B}=16.8$~mK, respectively, without
further adjustment of $T$ and $w_\mathrm{VDT}$, finding good
agreement with the experimental data.\\

Using the quantitative definitions (\ref{eq:p01}) and (\ref{eq:p99})
we can determine the optimal width of the optical tweezers for our
current experimental parameters, i. e., for the depth of the HDT of
$0.8$~mK and the atomic temperature of $T=60~\mathrm{\mu K}$. Using
(\ref{eq:p01}) we find the optimal depth of the VDT at
$U_\mathrm{VDT}^0/k_\mathrm{B}=0.5$~mK. The corresponding width of
the optical tweezers is calculated with (\ref{eq:p99}) to be
$2y_\mathrm{T}^\mathrm{opt}=2\cdot 11.7~\mathrm{\mu m}$, see
Tab.~\ref{table:tweezer_widths}.
\begin{table}[!t]
  \caption{\label{table:tweezer_widths}Width of the tweezers $2 y_\mathrm{T}^\mathrm{opt}$ for different parameters}
  \begin{tabular*}{1.0\textwidth}%
     {@{\extracolsep{\fill}}ccccccccr}\\

&$\frac{U_\mathrm{HDT}^0}{k_\mathrm{B}}~$(mK) & $\frac{U_\mathrm{VDT}^0}{k_\mathrm{B}}~$(mK) & $T~(\mathrm{\mu K})$& $\frac{w_\mathrm{HDT}}{\lambda_\mathrm{VDT}}$ & $\frac{w_\mathrm{VDT}}{\lambda_\mathrm{VDT}}$ &  $F_\mathrm{r}$ & $y_\mathrm{T}^\mathrm{opt}~(\mathrm{\mu m})$&\\
  \hline  % put a line under headers
  & \multicolumn{8}{l}{current experiment}\\
  1&0.8  & 0.51  & 60.0  & 19  & 9.8 &  0.015 & 11.7\\
  & \multicolumn{8}{l}{stronger focusing of optical tweezers}\\
  2&0.8  & 0.51  & 60.0  & 19  & 4.9 &  0.025& 5.9\\
  3&0.8  & 0.51  & 60.0  & 19  & 2.45 &  0.025 & 2.9\\
  & \multicolumn{8}{l}{lower atom temperature}\\
  4&0.8  & 0.017  & 1.0  & 19  & 4.9 &  0.482 & 2.9\\
  5&0.8  & 0.0088  & 0.084  & 19  & 2.45  & 0.920 & 0.5\\
%  & \multicolumn{8}{l}{stronger focusing of the HDT}\\
%  6&8.2  & 1.0  & 60.0  & 5  & 4.9 & 16.722 & 0.49768 & 2.87\\
%  %7&27  & 2.4  & 6  & 3.5  & 5 & 402 & 0.979 & 0.5\\
%  7&48.0  & 6.3  & 60.0  & 2.5  & 2.45 & 105.46 & 0.92035 & 0.5\\
%  6&0.8  & 0.1  & 1  & 5  & 2.5 & 104 & 0.919 & 0.5\\
%  7&8  & 1.0  & 10  & 2.5  & 2.5 & 104 & 0.919 & 0.5\\
%  8&16  & 1.3  & 5  & 4  & 4 & 259 & 0.968 & 0.5\\
    \hline  % put a line under headers
%\label{table:ch2_lenses}
  \end{tabular*}
 % \end{center}
 \end{table}

\subsubsection{Towards ultimate resolution}
Ultimate resolution of the optical tweezers is realized, if a single
potential well of the HDT is addressed only. Here, we use our model
in order to develop strategies for the reduction of the width of our
optical tweezers. It depends on the depth and waist of the VDT, of
the HDT, and on the temperature of the atoms. Experimentally,
variation of the depth of the traps is straight forwardly realized
by changing the power of the respective laser (up to 20~W for the
VDT laser and 1.2~W for each beam for the HDT laser). Changing the
waist size of the traps requires a new lens system, and lowering of
the atomic temperature could be achieved by e. g. Raman sideband
cooling technique \cite{Lee96}.

In the following analysis we ignore further experimental effects not
included in our model, e. g., drifts of the traps, fluctuations of
the trap depths, or heating in the traps, which
become relevant for ultimate precision.\\ \\
\textit{Universal extraction function} In order to introduce
dimensionless parameters, we rewrite (\ref{eq:uVDTmodified2}) in the
form
\begin{figure}
\centering
\includegraphics[width=8cm]{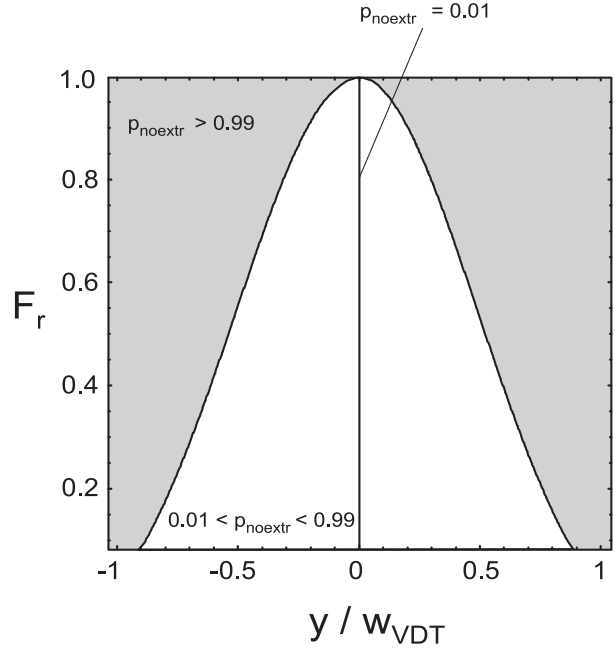}
\caption{\label{fig:pHDT_contourPlot} Contour plot for
$p_\mathrm{noextr}(y_\mathrm{T}^\mathrm{opt},F_\mathrm{r})=0.99$.
Outside the white area atoms will remain trapped in the HDT with
99~\% probability.}
\end{figure}
\begin{equation}
\frac{U_\mathrm{eff}(y)}{k_\mathrm{B}T}=s_\mathrm{T}\left(e^{-\frac{2y^2}{w_\mathrm{VDT}^2}}-
F_\mathrm{r} \right),
 \label{eq:invariants}
\end{equation}
where the normalized tweezers potential depth is
\begin{equation}
s_\mathrm{T}=\frac{U_\mathrm{VDT}^0}{k_\mathrm{B} T}
 \label{eq:sT}
\end{equation}
 and
\begin{equation}
F_\mathrm{r}=\frac{1}{\pi
\sqrt{e}}~\frac{U_\mathrm{HDT}^0}{U_\mathrm{VDT}^0}~\frac{\lambda_\mathrm{VDT}}{w_\mathrm{HDT}}
 \label{eq:Fr}
\end{equation}
is a relative measure of the forces exerted by the HDT ($\sim
U_\mathrm{HDT}^0/w_\mathrm{HDT}$) compared to VDT ($\sim
U_\mathrm{VDT}^0/\lambda_\mathrm{VDT}$). The condition of the
extraction (\ref{eq:uCondition}) translates into
$$F_\mathrm{r}<1.$$

%We determine, first, how the depths of both traps have to be changed
%in order to reach $y_\mathrm{T}=0.5~\mathrm{\mu m}$. We start the
%optimization from re-wring of Eq.~(\ref{eq:uVDTmodified2}) in a form
%appropriate for the analysis:
%\begin{equation}
%\frac{U(y)}{k_\mathrm{B}T}=a\left(e^{-\frac{2y^2}{w_\mathrm{VDT}^2}}-
%g \right),
% \label{eq:invariants}
%\end{equation}
%where $$a=\frac{U_\mathrm{VDT}^0}{k_\mathrm{B}
%T}~~~~\mathrm{and}~~~~
%g=\frac{1}{2\sqrt{e}}~\frac{U_\mathrm{HDT}^0}{U_\mathrm{VDT}^0}~\frac{\lambda_\mathrm{VDT}}{w_\mathrm{HDT}}.$$

The condition (\ref{eq:p01}) that the target atom is efficiently
extracted out of the HDT, is satisfied for
%\begin{equation}
%P_HDT(0)=0.01,
% \label{eq:condition_pHDT}
%\end{equation}
%see Eq.~(\ref{eq:pHDT}). This condition is satisfied when
\begin{equation}
s_\mathrm{T}=\frac{8.4}{1-F_\mathrm{r}}.
 \label{eq:condition_H}
\end{equation}
Now we rewrite (\ref{eq:p99}) in terms of the dimensionless
parameters $s_\mathrm{T}$ and $F_\mathrm{r}$ and use the
substitution \ref{eq:condition_H} to find the connection between the
optimal width of the optical tweezers $y_\mathrm{T}^\mathrm{opt}$
and the dimensionless parameter $F_\mathrm{r}$:
\begin{equation}
p_\mathrm{noextr}(y_\mathrm{T}^\mathrm{opt},F_\mathrm{r})=0.99,
 \label{eq:condition_p_noextr}
\end{equation}
which is plotted in figure~\ref{fig:pHDT_contourPlot}.

From this figure we can already infer two strategies for improved
resolution: the value of $F_\mathrm{r}$ should be about unity, and
the waist of the VDT should be as small as possible. $F_\mathrm{r}$
can be increased by increasing the depth of the HDT, by reducing
$w_\mathrm{HDT}$ and by lowering the temperature $T$ of the atoms,
see (\ref{eq:Fr}) and (\ref{eq:sT}).\\ \\
\textit{Examples of optical tweezers} In
Table~\ref{table:tweezer_widths} we have listed possible parameters
for the traps which would improve the extraction resolution. In line
1 we have optimized $U_\mathrm{VDT}^0$ for our experimental
parameters. In lines 2-3 we project parameters for improved
resolution by changing the focus of the VDT, in lines 4 and 5 the
effect of lower temperatures is shown (about 1 $\mu$K and 0.1 $\mu$K
which can be obtained with Raman cooling \cite{Perrin98} and quantum
degenerate gases).

\subsection{Insertion of an atom}
\label{subs:insertion}

\begin{figure}
\centering
\includegraphics[width=9cm]{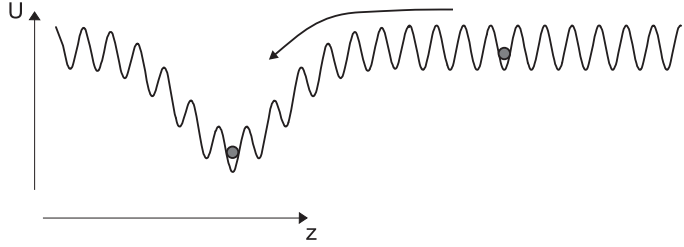}
\caption{\label{fig:insertionAxial} Axial insertion. An atom trapped
in one of the potential wells of the standing wave of the VDT, is
inserted into the Gaussian potential well of the HDT by axially
moving the VDT along the $z$-direction.}
\end{figure}

\begin{figure}
\centering
\includegraphics[width=14cm]{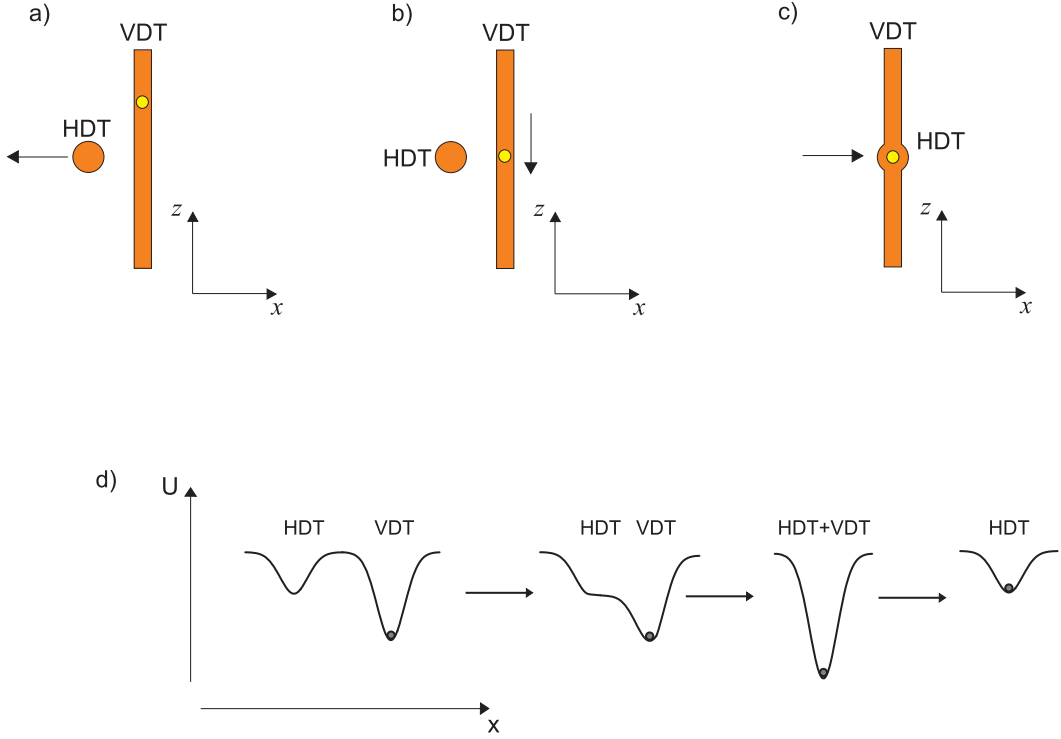}
\caption{\label{fig:radialInsertionScheme}Radial insertion of an
atom. a) An atom in the VDT after the extraction. The traps are
separated by displacing the HDT along the $x$-direction. b) The atom
in the VDT is transported to the $z$-position of the HDT. c) The
traps are merged by moving the HDT along the $x$-direction towards
the VDT. d) Evolution of the radial potentials of the traps along
the $x$-axis for steps b) and  c).}
\end{figure}

%Aim: Introduce the physical process necessary for the insertion.
%Introduce the axial, see Fig.~\ref{fig:insertion}a and the radial
%methods, see Fig.~\ref{fig:radialInsertionScheme} and
%Fig.~\ref{fig:insertion}b.\\
After extraction, the atom is trapped in the potential of the VDT.
In order to re-insert the atom into a potential well of the HDT, the
VDT potential is merged with the HDT and finally switched off. There
are two alternative methods to insert an atom back into the HDT:
``axial insertion'' and ``radial insertion''.\\
\\
 \textit{Axial insertion}
In this case, the process of extraction of an atom is simply
reversed: the VDT axially transports the atom to the axis of the
HDT, see figure~\ref{fig:insertionAxial}, and then the VDT is
adiabatically switched off, leaving the atom in the HDT. The whole
process of axial insertion takes about $70~\mathrm{ms}$.\\
\\
\textit{ Radial insertion} For radial insertion, the two traps are
first radially separated by displacing the axis of the HDT in the
positive $x$-direction, see
figure~\ref{fig:radialInsertionScheme}(a). Then the atom in the VDT
is transported downwards to the vertical position of the horizontal
trap, see figure~\ref{fig:radialInsertionScheme}(b).
Along the $x$-axis, the atom in this configuration is confined in
the Gaussian-shaped radial potential of the VDT. In the next step,
the VDT is then merged with the Gaussian-shaped radial potential of
the HDT by moving the HDT radially towards the $x$-position of the
VDT, see figure~\ref{fig:radialInsertionScheme}(c). In the final
step the VDT is adiabatically switched off, which releases the atom
to the HDT, see figure~\ref{fig:radialInsertionScheme}(d). The
process of radial
re-insertion takes about $210~\mathrm{ms}$.\\

A radical difference of the two alternative insertion methods occurs
if the HDT already holds an atom within the width of the optical
tweezers. During axial insertion the VDT exerts the same forces as
during the extraction of an atom. Therefore, the achievable final
distance between two atoms in the HDT is limited to the width of the
optical tweezers, because atoms within the extraction region will be
extracted downwards by the VDT during the re-insertion. In contrast,
if the two traps are merged radially, the VDT does not exert any
forces which could push an atom out of the HDT. Consequently, for
radial insertion there are no limitations on the final separations.
In particular, the final distance between two atoms could be set to
zero. In this way, two atoms could be joined in a single
micropotential of the standing wave of the HDT. A disadvantage of
the radial insertion is additional heating of the atom in the HDT,
see Sec.~\ref{subs:insertion}.

The ultimate goal of insertion is to reliably place an atom into a
given micropotential of the HDT, for instance an integer number of
potential wells away from the neighboring atom, but without
influencing it.

\subsubsection{Insertion precision}
\label{subsubs:insertPrecision} There are two independent effects
influencing the precision of the insertion, i. e., how accurately an
atom can be placed at a desired position of the HDT: the thermal
motion in the VDT and the position fluctuations of the VDT relative
to the HDT. Both of them equally affect the axial and the radial
insertion. Therefore, the following theory applies to both methods
of insertion.\\
\\
 \textit{Thermal distribution in the VDT}
Before contact with the HDT atoms in the VDT are distributed
thermally along the $y$-direction (the axis of the HDT) in an
approximately harmonic potential with oscillation frequency
$\Omega_\mathrm{rad}=\sqrt{\frac{4U^0_\mathrm{VDT}}{m
w_\mathrm{VDT}^2}}$. It is known that the distribution in this case
is a Gaussian,
\begin{equation}
%p_\mathrm{T}(y)=\int_0^\infty
%p_{\varepsilon}(y)f_1(\varepsilon,T)d\varepsilon=\sqrt{\frac{m\Omega^2}{2\pi
%k_\mathrm{B}T}}e^{-m\Omega^2y^2/(2 k_\mathrm{B} T)},
p_\mathrm{T}(y)=\sqrt{\frac{m\Omega_\mathrm{rad}^2}{2\pi
k_\mathrm{B}T}}e^{-m\Omega_\mathrm{rad}^2y^2/(2 k_\mathrm{B} T)}.
 \label{eq:spatial_distrbution}
\end{equation}
The width of this distribution
\begin{equation}
\delta
y_\mathrm{therm}=\frac{w_\mathrm{VDT}}{2}\sqrt{\frac{k_\mathrm{B}
T}{U_\mathrm{VDT}^0}}
 \label{eq:insertion_precisionThermal}
\end{equation}
can also be expressed in terms of the VDT waist radius and the
$s_\mathrm{T}$-parameter, combining $U_\mathrm{VDT}^0$ and the
temperature, see (\ref{eq:sT}):
\begin{equation}
\delta y_\mathrm{therm}=\frac{w_\mathrm{VDT}}{2s_\mathrm{T}^{1/2}}~.\\
\\
 \label{eq:insertion_precision_SParameter}
\end{equation}
\textit{Spatial fluctuations of the VDT} Since the VDT and the HDT
laser beams are guided by independent mechanical setups, their
relative position is subject to radial and axial fluctuations. In
our model these fluctuations are taken into account by $\delta
y_\mathrm{fluct}$ representing the rms-amplitude of the
fluctuations of the VDT axis.\\ \\

\begin{figure}
\centering
\includegraphics[width=7cm]{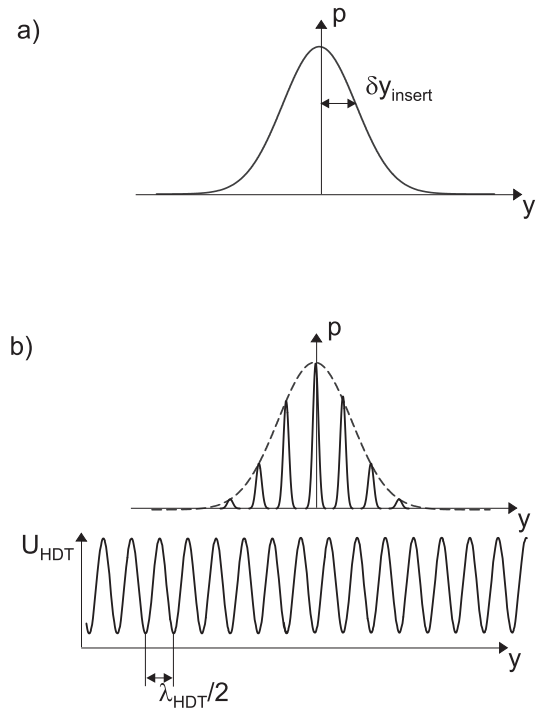}
\caption{\label{fig:projection}Spatial distribution $p(y)$ with the
width $y_\mathrm{insert}$ is projected onto the standing wave of the
HDT as both traps are merged (a). This distribution after the
projection (b). The width of the envelope remains almost unchanged,
but the probability is spatially modulated with the periodicity of
the HDT.}
\end{figure}

For our typical experimental parameters, the width of the thermal
distribution is on the order of $0.5~\mathrm{\mu m}$, and $\delta
y_\mathrm{fluct}$ is about $0.5~\mathrm{\mu m}$. Assuming these
fluctuations are Gaussian distributed the rms-amplitude of the
combined fluctuation is:
\begin{equation}
\delta y_\mathrm{insert}=\sqrt{\delta y_\mathrm{therm}^2+\delta
y_\mathrm{fluct}^2}~.
 \label{eq:insertion_precision}
\end{equation}
The value of $\delta y_\mathrm{insert}\approx 0.7~\mathrm{\mu m}$ is
the width of the distribution of the probability to find an atom
along the HDT axis $p(y)$. Since this value is larger than the size
of one HDT micropotential, the distribution extends over several
potential wells, see figure~\ref{fig:projection}(a).\\
\\
\textit{Insertion into HDT micropotentials by ``projection''} In the
last step of the insertion, the traps are merged and the VDT is
finally switched off. Due to the periodicity of the HDT, the
distribution $p(y)$ is changed: its envelope reflects the width of
the original distribution before the traps were merged, but under
this envelope the distribution is now modulated with the periodicity
of the standing wave of the HDT, see figure~\ref{fig:projection}(b).
In harmonic approximation, the distribution in each micropotential
is described again by a Gaussian of width
$$\delta y_\mathrm{micropot}=\frac{\lambda_\mathrm{HDT}}{2\sqrt{2}\pi}\sqrt{\frac{k_\mathrm{B} T}{U_\mathrm{HDT}^0}},$$
where $T$ is the temperature.

It is clear that the insertion precision will be improved by better
localization of the atoms, i. e., with lower atomic temperature and
deeper VDT potentials. Ultimately, for $\delta y_\mathrm{insert}\ll
\lambda_\mathrm{HDT}/2$ the final distribution will be concentrated
into a single micropotential. This limit corresponds to ``perfect''
insertion.

\subsubsection{Experimental studies of the insertion precision}
We have carried out a series of measurements in order to
experimentally study the dependence of the insertion precision on
atomic temperature and on the depth of the VDT predicted by the
above model. For this purpose we have loaded atoms into the HDT and
extracted the rightmost atom with the VDT, the rest of the atoms was
expelled out of the HDT by switching it off for 30~ms. The events
with no or too closely spaced atoms were discarded. The atom in the
VDT was then cooled with optical molasses, and placed back into the
HDT using the method of axial re-insertion. The final positions of
the inserted atoms were determined and the standard deviation
$\delta y_\mathrm{insert}$ was calculated. For $\delta
y_\mathrm{insert}\gtrsim 0.5 \lambda_\mathrm{HDT}$ we can neglect
the discretization of the positions due to the periodic structure of
the HDT \cite{Falk84}.

\begin{figure}
\centering
\includegraphics[width=15.5cm]{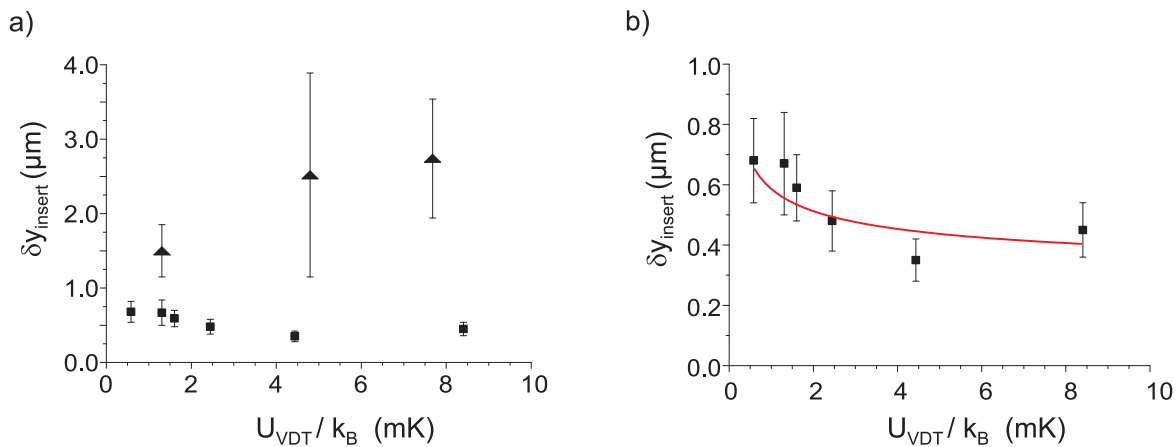}
\caption{\label{fig:reinsertionMeasurements}a) Insertion precision
as a function of the depth of the VDT. The triangular points present
the insertion precision for the experimental sequence without
cooling of the atoms before re-insertion. The rectangular points are
the results of the insertion precision with the cooling step. b)Zoom
of a) showing the data for precooled atoms. The solid line is the
fit of (\ref{eq:reinsertionFitFunction}) to the experimental data.
In this experiment we have used the axial insertion method. Every
point corresponds to about 25 repetitions of the experiment.}
\end{figure}
\subsubsection{Analysis}
Figure~\ref{fig:reinsertionMeasurements}(a) shows the dependence of
the measured insertion accuracy $\delta y_\mathrm{insert}$ on the
depth of the VDT. The corresponding graph for the radial insertion,
see figure~\ref{fig:reinsertionOptimalPower}(b), shows comparable insertion precision as expected.\\
\\
\textit{Temperature of the atoms} The temperature was varied by
performing the measurement with (squares) and without the cooling
step in the VDT (triangles). The huge difference in $\delta
y_\mathrm{insert}$ qualitatively demonstrates the temperature
dependency of the insertion precision and points out the importance
of the cooling step. The extraction process itself can heat up the
atom if it is initially not located on the VDT axis: the atom
remains at its $y$-position until it is released from its HDT
potential well and starts to oscillate radially in the VDT.
Therefore we have to cool the atom before insertion to achieve a
good insertion precision.\\
\\
\textit{Depth of the VDT} In order to insert the atoms at different
$U_\mathrm{VDT}$, we have first extracted and cooled the atoms at a
fixed depth $U_\mathrm{VDT}^\mathrm{extr}$ to insure constant
cooling parameters, and then adiabatically changed the depth to
$U_\mathrm{VDT}$. During this ramp, the temperature of the atoms
adiabatically changes to $T=T_0\sqrt{U_\mathrm{VDT} /
U_\mathrm{VDT}^\mathrm{extr}}$ ~\cite{Alt03}. Using this temperature
in (\ref{eq:insertion_precision}) and
(\ref{eq:insertion_precisionThermal}) we obtain the expected
insertion precision
\begin{equation}
\delta
y_\mathrm{exp}(U_\mathrm{VDT})=\sqrt{\frac{b^2}{\sqrt{U_\mathrm{VDT}/k_\mathrm{B}}}+\delta
y_\mathrm{fluct}^2}.
 \label{eq:reinsertionFitFunction}
\end{equation}
The parameter $b=\frac{w_\mathrm{VDT}}{2}
\frac{\sqrt{T_0}}{\sqrt[4]{U_\mathrm{VDT}^\mathrm{extr}/k_\mathrm{B}}}$
is a combination of the waist of the VDT, of the trap depth where
the atom was cooled and the temperature. We have independently
determined $\delta y_\mathrm{fluct}=0.26(\pm0.03)~\mathrm{\mu m}$ by
measuring the position of an atom in the VDT over the typical
duration of an experimental run (200~sec). Equation
(\ref{eq:reinsertionFitFunction}) was then fitted to the
experimental data with the fit parameter $b$, see
figure~\ref{fig:reinsertionMeasurements}(b), yielding $b=0.52(\pm
0.05)~\mathrm{\mu m}~\mathrm{mK}^{1/4}$.

\begin{figure}
\centering
\includegraphics[width=10cm]{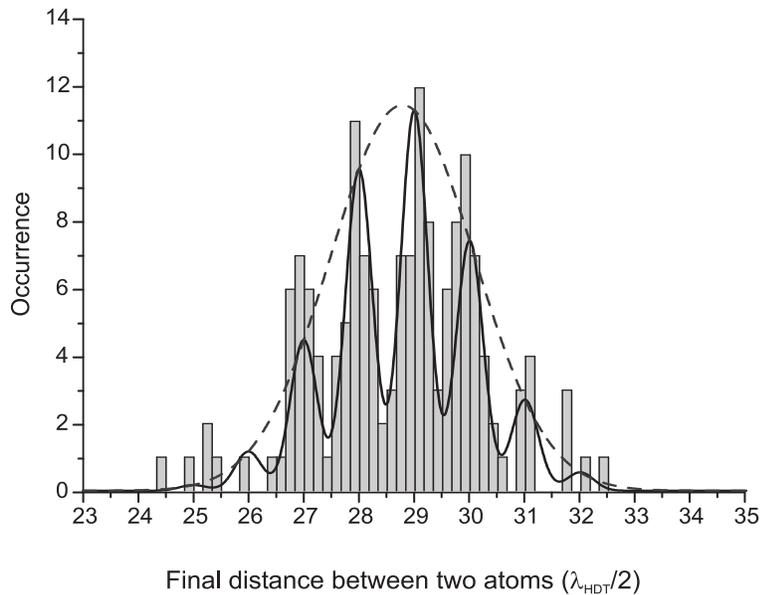}
\caption{\label{fig:finalDistancesDiscretization}Distribution of the
separations between two simultaneously trapped atoms after setting
their distance to $15~\mathrm{\mu m}$ with our distance-control
operation \cite{Miroshnychenko06}. The histogram clearly shows that
the distances are integer multiples of the standing wave period
$\lambda_\mathrm{HDT}$/2, and extending over only about 4 standing
wave potential wells. The solid line is a theoretical fit with a
Gaussian envelope (dashed line) centered at $15.31~\mu m$ and having
a $1/\sqrt{e}$-halfwidth of $0.71~\mu m$. The narrow peaks under
this envelope have a $1/\sqrt{e}$-halfwidth of $0.130~\mu m$,
corresponding to the precision of an individual distance measurement
\cite{Dotsenko05}.}
\end{figure}

Using
$U_\mathrm{VDT}^\mathrm{extr}/k_\mathrm{B}=1.6(\pm0.3)~\mathrm{mK}$
and the independently measured
$w_\mathrm{VDT}=10.1(\pm1.4)~\mathrm{\mu m}$ we calculate the
corresponding temperature of the atom in the VDT after the cooling
step to be $T_0=13(\pm 4)~\mu K$. This temperature is smaller than
the typical temperatures measured in the HDT. The difference between
these values could be explained by the fact that the multi mode
operation of our HDT laser impairs the cooling process in the HDT
\cite{Schrader04}, or by systematic errors due to approximations in
our model.\\
\\
\textit{Periodicity of the HDT} Until now we have analyzed the
insertion precision in the frame of reference of our ICCD camera.
Much more important is the insertion precision relative to a fixed
point in the HDT, e. g., another atom. Here we have prepared a pair
of atoms with a fixed separation using our distance-control
operation, Sec.~\ref{subs:repositioning_procedure}. Instead of final
positions we now measure final distances, see
figure~\ref{fig:finalDistancesDiscretization}. Here the periodicity
of the HDT is clearly visible as it is expected from
figure~\ref{fig:projection}(b). In the distance measurement the
random axial shot--to--shot fluctuations of the HDT standing wave
pattern cancel out, whereas in position measurements relative to the
ICCD this modulation is smeared out.

The insertion precision relative to a second atom in the HDT is on
the same order of magnitude as $\delta y_\mathrm{insert}$ measured
in the previous section. This allows us to set a distance between
two atoms with an accuracy corresponding to about four potential
wells of the HDT.

\subsection{Insertion induced heating}
\label{subs:insertion_heting} As discussed in
section~\ref{subs:insertion}, the method of radial insertion allows
us to place an atom arbitrary close to other atoms in the HDT. It
turns out that during this insertion an atom in the HDT at the
position of the VDT is heated up. This heating effect limits the
usable depth of the VDT, such that a compromise between a high
precision of insertion and tolerable heating is required.

\subsubsection{Adiabatic model}
\begin{figure}
\centering
\includegraphics[width=7cm]{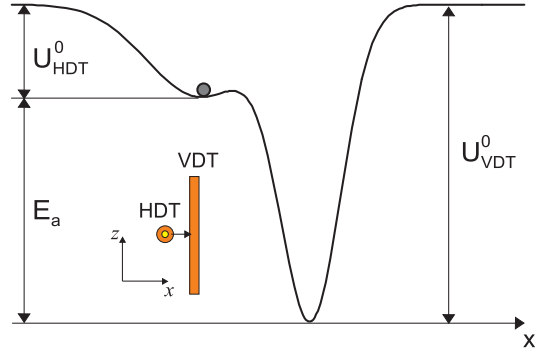}
\caption{\label{fig:reinsertionHeating}Radial potential of the two
traps. During the radial merging of the traps, an atom in the radial
potential of the HDT has a potential energy $E_\mathrm{a}$ relative
to the bottom of the potential well of the VDT. The inset shows the
respective geometry of the traps.}
\end{figure}

Consider an atom trapped in the HDT at the $y$-position of the VDT,
when the traps are axially separated along the $x$-direction. Along
this direction the potential is the sum of two Gaussians, i. e., the
radial potentials of the two traps. Just before the two traps are
merged, the potential shape is shown in
figure~\ref{fig:reinsertionHeating}. For $k_\mathrm{B}T\ll
U_\mathrm{HDT}^0$, the atom stays near the bottom of the HDT
potential until it falls down into the VDT potential. With respect
to the bottom of this potential, it has an energy of approximately
$E_\mathrm{a}\approx U_\mathrm{VDT}^0-U_\mathrm{HDT}^0$.
Adiabatically switching off of the VDT causes the atom to be
adiabatically cooled to the final atomic energy \cite{Alt03}:
\begin{equation}
E_\mathrm{a}^\mathrm{final} \approx
E_\mathrm{a}\sqrt{\frac{U_\mathrm{HDT}^0}{
U_\mathrm{HDT}^0+U_\mathrm{VDT}^0}},
\end{equation}
where the difference between $w_\mathrm{VDT}$ and $w_\mathrm{HDT}$
has been neglected. The condition for the atom to remain trapped in
the HDT is $E_\mathrm{a}^\mathrm{final}<U_\mathrm{HDT}^0$, yielding
an upper limit for the depth of the VDT
\begin{equation}
U_\mathrm{VDT}^0\lesssim 3 U_\mathrm{HDT}^0,
 \label{eq:reinsertionHeating}
\end{equation}
otherwise the atom will be lost.
\begin{figure}
\centering
\includegraphics[width=7.5cm]{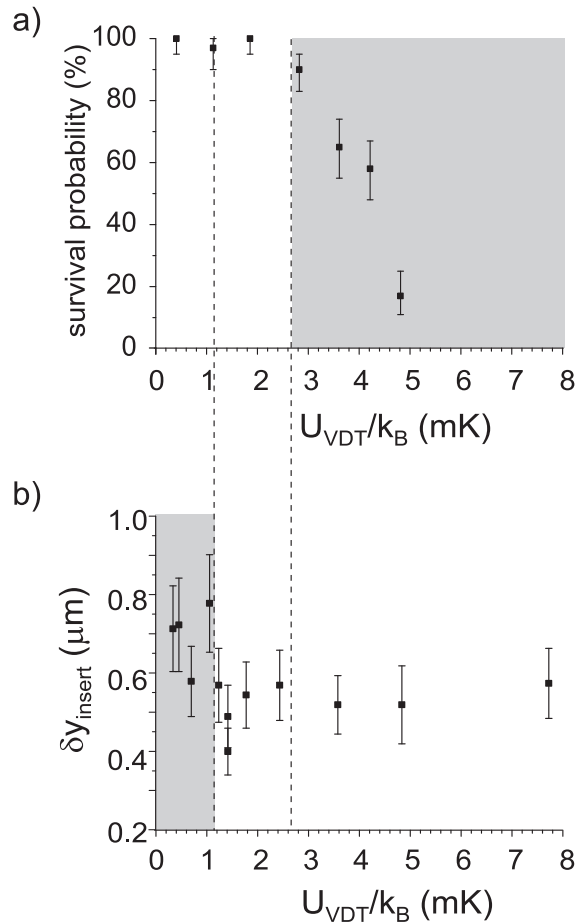}
\caption{\label{fig:reinsertionOptimalPower}Optimal depth of the
VDT. a) The survival probability of an atom in the HDT as a function
of the depth of the VDT. Every point is the result of about 35
repetitions of the experiment. b) The insertion precision as a
function of the depth of the VDT. Every point corresponds to about
35 radial insertions with a single atom. The shaded areas show the
experimentally unfavorable ranges of the depth of the VDT.}
\end{figure}
\subsubsection{Measurement of the heating effect} One atom on average was loaded into the
HDT and transported to the $y$-position of the VDT axis. For the
third step, the HDT with the atom was transported in the
$x$-direction, and the VDT was switched on. At the fourth step, the
atom was transported back towards the VDT as it would occur during
the radial insertion. Thereafter, the VDT was adiabatically switched
off. The final image reveals then the presence or loss of the atom
in the HDT. Figure~\ref{fig:reinsertionOptimalPower}(a) shows the
survival probability of the atom after this manipulation.

\subsubsection{Analysis}
The experimental data in
figure~\ref{fig:reinsertionOptimalPower}(a), show that starting from
a VDT depth of about 2.5~mK, the atoms in the HDT were heated up and
lost during the radial insertion procedure. Since the depth of the
HDT for this experiment was 0.8~mK, the condition
(\ref{eq:reinsertionHeating}) results in $U_\mathrm{VDT}^0<2.4$~mK,
which reasonably agrees with the experimentally observed value.

At the same time, the lower limit on the depth of the VDT is
dictated by the insertion precision, which deteriorates with the
reduction of the VDT depth.
Figure~\ref{fig:reinsertionOptimalPower}(b) shows the insertion
precision, measured for the same depth of the HDT using the radial
insertion  method. For VDT depths below 1.2~mK the insertion
precision is dominated by the thermal component and starts to
deteriorate, see (\ref{eq:insertion_precisionThermal}).

The non shaded regions in figure~\ref{fig:reinsertionOptimalPower}
indicate the range of experimentally useful depths of the VDT. For
our typical experimental parameters there is a non empty overlap of
these regions. It can be further enlarged by increasing the depth of
the HDT according to (\ref{eq:reinsertionHeating}).

%On the other side, if the
%radial drifts and fluctuations of the VDT were reduced, the
%thermal-atom contribution to the insertion precision, see
%Eq.~(\ref{eq:insertion_precision}) and
%Eq.~(\ref{eq:insertion_precisionThermal}),
%
%there would be necessary to reduce the thermal-atom contribution to
%the insertion precision, see Eq.~(\ref{eq:insertion_precision}) and
%Eq.~(\ref{eq:insertion_precisionThermal}). This can be realized by
%increasing the depth of the VDT.\\

\subsection{Conclusion}
Using optical tweezers we have repositioned individual atoms inside
a standing wave optical dipole trap with a precision on the order of
the periodicity of the trap. Regular string containing 2 to 7 atoms
have been prepared atom--by--atom. We have modeled and
experimentally analyzed the processes of extraction and insertion of
a single atom in detail. We have identified the main limiting
factors and proposed strategies for future improvements.

We demonstrate two methods of insertion, one of which has no
limitation on the final distances between the atoms after the
re-insertion. It can therefore be made as small as zero, i. e.,
placing two atoms into the same potential well of the standing wave.
We have found suitable parameters of the traps, which allow us to
perform the re-insertion efficiently and with high precision.

\end{document}